\documentclass[12pt]{iopart}

\usepackage{amsfonts,graphicx,iopams}

\newcommand{\hs}[1]{\hspace{#1 cm}}
\newcommand{\betat}{\tilde{\beta}}
\newcommand{\mut}{\tilde{\mu}}
\newcommand{\omegat}{\tilde{\omega}}
\newcommand{\rhot}{\tilde{\rho}}
\newcommand{\gammat}{\tilde{\gamma}}
\newcommand{\Pt}{\tilde{P}}
\newcommand{\epsmin}{\epsilon_-(\alpha)}
\newcommand{\epsplus}{\epsilon_+(\alpha)}
\newcommand{\epm}{{\epsilon_-}}
\newcommand{\epp}{{\epsilon_+}}

\begin{document}
\title{The entropy of dense non-commutative fermion gases}  
\date{\today}
\author{J N Kriel$^1$ and F G Scholtz$^{1,2}$}
\address{$^1$ Institute of Theoretical Physics, University of Stellenbosch, Stellenbosch 7600, South Africa}
\address{$^2$ National Institute for Theoretical Physics (NITheP), 7600 Stellenbosch, South Africa}

\begin{abstract}
We investigate the properties of two- and three-dimensional non-commutative fermion gases with fixed total z-component of angular momentum, $J_z$, and at high density for the simplest form of non-commutativity involving constant spatial commutators.  Analytic expressions for the entropy and pressure are found.  The entropy exhibits non-extensive behaviour while the pressure reveals the presence of incompressibility in two, but not in three dimensions.  
Remarkably, for two-dimensional systems close to the incompressible density, the entropy is proportional to the square root of the system size, i.e., for such systems the number of microscopic degrees of freedom is determined by the circumference, rather than the area (size) of the system. The absence of incompressibility in three dimensions, and subsequently also the absence of a scaling law for the entropy analogous to the one found in two dimensions, is attributed to the form of the non-commutativity used here, the breaking of the rotational symmetry it implies and the subsequent constraint on $J_z$, rather than the angular momentum $J$. 
Restoring the rotational symmetry while constraining the total angular momentum $J$ seems to be crucial for incompressibility in three dimensions. We briefly discuss ways in which this may be done and point out possible obstacles.   
\end{abstract}

\section{Introduction}

The realization that gravitational instabilities make the absolute localization of events impossible, which in turn may be captured through a non-commutative space-time structure \cite{dop}, has led to considerable interest in the formulation of non-commutative quantum field theories \cite{doug}, quantum mechanics \cite{scholtz2009} and the possible physical consequences of non-commutativity \cite{li,mendes,bem,khan,chai,chair,lia,chai1,gar,alex}.  

One aspect of non-commutative systems that has received very little attention in the literature is the thermodynamics.  The reason for this is probably three-fold.  Firstly, to discuss the thermodynamics one has to consider systems confined to a finite volume, e.g. a box, to investigate the dependence of the thermodynamics on the system size.  The introduction of boundaries for non-commutative systems is notoriously difficult and has only been done successfully for a two-dimensional well \cite{scholtz2007}.  Secondly, apart from the example just mentioned, there are no other non-trivial (apart from the harmonic oscillator) systems for which the energy spectrum is known, even numerically, which makes the computation of the thermodynamics intractable. Here it should be mentioned that the spectra of a number of non-commutative systems have been computed perturbatively to leading order in the non-commutative parameter \cite{li,mendes,bem,khan,chai,chair,lia,chai1,gar,alex}.  However, as will become clear later, this is not sufficient if one wants to understand the physically interesting high temperature or high density limits, which show strong deviations from commutative behaviour. Thirdly, the issue of the breaking of rotational symmetry in three dimensions creates a conceptual barrier.  One either has to accept it, seek modified commutation relations that respect the symmetry or one can restore the symmetry through twisting \cite{chai1,wess1,asch,bal,chak1}.  If the latter route is chosen, it also becomes necessary to twist the statistics, but there seems to be disagreement on whether this should be done in the sense of \cite{bal,chak1}, which results in modified statistical correlations \cite{chak1}, or in the braided sense of \cite{wess,fiore}.  

The first attempt to understand the thermodynamics of such systems was made in \cite{scholtz2008}.  In this case the difficulties above were avoided by considering particles confined in a two-dimensional infinite well for which the spectrum is known \cite{scholtz2007}.  The issues of rotational invariance and the twisting of statistics are also not present.  However, the computation could still only be carried out numerically and thus the thermodynamic limit could not be probed reliably. The results of these calculations were quite striking and showed strong deviations from commutative behaviour at high densities and temperatures. The most important feature is the appearance of incompressibility at high densities for fixed angular momentum.  This reflects the effective excluded volume implied by the non-commutative length scale. Under these conditions several other thermodynamic features also deviate strongly from the commutative case and, in particular, the entropy is no longer extensive.  

The surprising nature of these results provide a strong motivation for a better understanding, both in terms of a more reliable analytic treatment that can probe the thermodynamic and high density limits more efficiently and the generalization to three dimensions. With regard to the latter, the issue of symmetry breaking still prevails.  Here we simply accept the breaking of rotational symmetry and investigate the consequences without any attempt to restore the symmetry through modified commutation relations or twisting. The naive expectation is that the breaking of rotational symmetry should be irrelevant in the thermodynamic limit.  The essential reasoning behind this is that the shape of the container we use to study the system, be it a sphere or a cylinder, is immaterial in the thermodynamic limit and only shows up in finite size corrections.  However, it turns out that this expectation is only realized in the low density limit, while it fails in the high density limit. The rest of the high density thermodynamics also do not reflect the desired features and in particular incompressibility. The expectation, therefore, is that symmetry breaking matters at high density and that the restoration thereof will be a key element for the understanding of the high density thermodynamics.  
The possible scenarios for restoration of the rotational symmetry and the possible problems associated with them are discussed later.  

This sets the scene for the present paper, which aims at understanding the thermodynamics of non-commutative fermion gases in two and three dimensions analytically, particularly at high densities and low temperature. The paper is organized as follows. Section \ref{twodim} reviews and generalizes the results of \cite{scholtz2008} for the two-dimensional gas.  The focus here is on the analytic treatment of the high density limit. Section \ref{critical} focuses on the two-dimensional thermodynamics close to the incompressible point and it is shown that the entropy exhibits a square root dependence on the system size. Section  \ref{threedim} generalizes these results to three dimensions and discusses the issues around the restoration of the rotational symmetry. The paper is concluded in section \ref{concl} with a summary and outlook.        

\section{The two-dimensional non-commutative fermion gas}
\label{twodim}
\subsection{Definitions}
We begin by summarizing the key results of \cite{scholtz2007,scholtz2008}. Consider a system of $N$ spinless particles of mass $m_0$ which are confined to a disc with area $A=\pi R^2$. The particle coordinates satisfy $[x,y]=i\theta$ ($\theta>0$) and we set $R^2=\theta (2M+1)$ with $M$ a positive integer. The latter serves as a dimensionless measure of the system size. The single particle energies are then $E_{r,m}=\hbar^2 x_{r,m}/(\theta m_0)$ with $x_{r,m}$ the zeros of the Laguerre polynomial $L_{M+1}^m(x)$ ($m\geq0$) or $L_{M+m+1}^{|m|}(x)$  ($-M\leq m<0$). Here $m$ is an angular momentum label while $r$ labels the $M+1$ ($m\geq0$) or $M+m+1$ ($-M\leq m<0$) states within a specific angular momentum sector. Note that $m$ is bounded from below by $-M$ and so there are only a finite number of negative angular momentum states.\\ 

The $q$-potential of the grand canonical ensemble is 
\begin{equation}
	q(A,\beta,\mu,\omega)=\sum_{m=-M}^{\infty}\sum_r\log[1+e^{-\beta (E_{r,m}-\mu-\hbar\omega m)}].
	\label{qpotoriginal}
\end{equation}
We set $E_0=\hbar^2/(\theta m_0)$ and introduce the dimensionless parameters $\betat=E_0\beta$, $\mut=\mu/E_0$ and $\omegat=\hbar\omega/E_0$ in terms of which the $q$-potential becomes
\begin{equation}
	q(M,\betat,\mut,\omegat)=\sum_{m=-M}^{\infty}\sum_r\log[1+e^{-\betat (x_{r,m}-\mut-\omegat m)}].
	\label{qpotdimless}
\end{equation}
The central thermodynamic quantities can be calculated from $q$ using
\begin{equation}
	N=\frac{1}{\betat}\frac{\partial q}{\partial \mut},\hs{2} \frac{L}{\hbar}=\frac{1}{\betat}\frac{\partial q}{\partial \omegat},\hs{2} \frac{S}{k}=q-\betat\frac{\partial q}{\partial \betat}
	\label{thermorelations}
\end{equation}
and we also define dimensionless measures of the density and pressure by $\rhot=N/(2M+1)$ and $\Pt=q/(\betat(2M+1))$.\\

In the thermodynamic limit \eref{qpotdimless} can be replaced by
\begin{equation}
	q(M,\betat,\mut,\omegat)=\int_{-M}^{\infty}{\rm d}m\int_{x_-(m)}^{x_+(m)}{\rm d}x\ D(x,m)\log[1+e^{-\betat (x-\mut-\omegat m)}]
	\label{qpotdimlessintegral}
\end{equation}
where $D(x,m)$ is the asymptotic density of the zeros of the Laguerre polynomial $L_{M+1}^m(x)$ ($m\geq0$) or $L_{M+m+1}^{|m|}(x)$  ($-M\leq m<0$) in the limit where $M$ and $m$ tend to infinity in a fixed ratio. It is known that $D(x,m)$ is given by \cite{dette1995}
\begin{equation}
	D(x,m)=\frac{\sqrt{4 M x-(m-x)^2}}{2 \pi  x}
	\label{laguerrezerodensity}
\end{equation}
for $x$ in the interval on which the argument of the radical is non-negative.
\begin{figure}[t]
\begin{tabular}{cc}
\includegraphics[width=7.5cm]{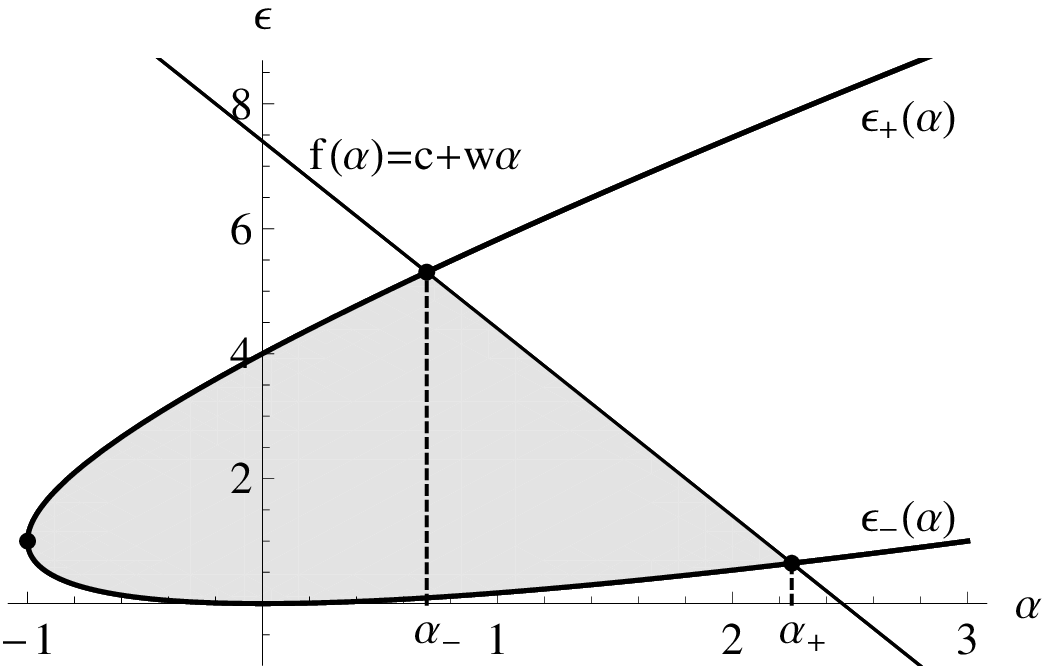} & \includegraphics[width=7.5cm]{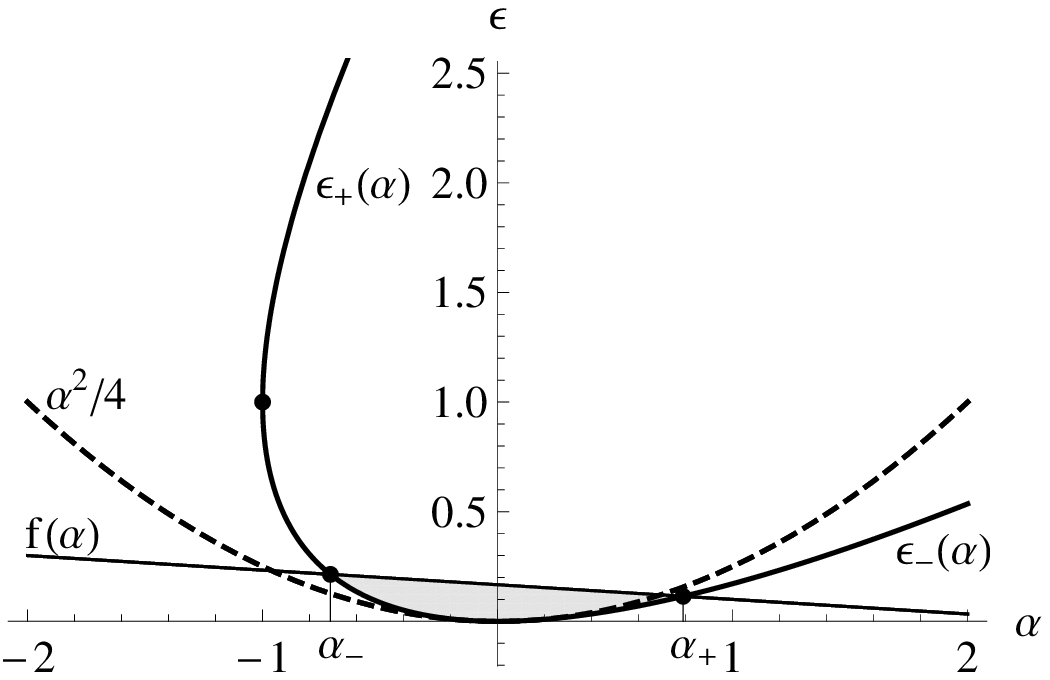}\\
(a) & (b)
\end{tabular}
  \caption{The curves $\epsilon_{\pm}(\alpha)$ define the range of the rescaled single particle energy $\epsilon$ for each value of $\alpha$. These two curves meet at the point $(-1,1)$ which is indicated by a dot. (a) The line $f(\alpha)$, shown here for $c=7.4$ and $w=-3$, defines an effective chemical potential within each angular momentum sector. The points $\alpha_-$ and $\alpha_+$ denote the intersections of $f(\alpha)$ with the $\epsilon_{\pm}(\alpha)$ curves. At low temperatures the integrand in \eref{qpot} is appreciable only within the shaded region. (b) An illustration of the low density case with small values of $c$ and $w$. }\label{domainplot}
\end{figure}
\subsection{The low-density limit}
To analyse the integral in \eref{qpotdimlessintegral} we first transform to the rescaled parameters $\epsilon=x/M$, $\alpha=m/M$, $b=M\betat$, $c=\mut/M$ and $w=\omegat$. The $q$-potential then reads
\begin{equation}
	q(M,b,c,w)=M^2\int_{-1}^{\infty}{\rm d}\alpha\int_{\epsmin}^{\epsplus}{\rm d}\epsilon\ d(\epsilon,\alpha)\log[1+e^{-b (\epsilon-f(\alpha))}] \label{qpot}
\end{equation}
where $f(\alpha)=c+w\alpha$ and
\begin{equation}
	d(\epsilon,\alpha)=D(M\epsilon,M\alpha)=\frac{\sqrt{4 \epsilon -(\alpha -\epsilon )^2}}{2 \pi  \epsilon }.
\end{equation}
The latter is defined for $\epsilon\in[\epsmin,\epsplus]$ with 
\begin{equation}
	\epsilon_{\pm}(\alpha)=\alpha+2 \pm 2 \sqrt{\alpha +1}.
\end{equation}
The form of the integrand in \eref{qpot} suggests that $f(\alpha)$ acts as an effective chemical potential within the $m=\alpha M$ angular momentum sector. At low temperatures almost the entire contribution to the integral therefore comes from the region defined by $\epsmin\leq\epsilon\leq\min[f(\alpha),\epsplus]$. This is illustrated in figure \ref{domainplot} for particular choices of parameters.\\

Next we illustrate how at low densities non-commutative effects become unimportant and the physics reduces to that of a standard commutative fermion gas. 
 
Consider a process in which the system size $M$ is increased at a fixed particle density and angular momentum $L=0$. Since the chemical potential $\mut$ is intensive we expect $c=\mut/M$ to approach zero as $M$ tends towards infinity. From figure \ref{domainplot} (b) it is clear that to produce a zero total angular momentum $w$ must itself be approximately zero with the intersections $\alpha_-$ and $\alpha_+$ located roughly symmetrically around $\alpha=0$ at $\alpha_\pm\approx \pm2\sqrt{c}$. At low temperatures almost the entire contribution to \eref{qpot} will therefore come from a region where $|\alpha|\sim\sqrt{c}$ and $\epsilon\sim c$. This implies that in \eref{qpot} we may approximate $d(\epsilon,\alpha)$ by
\begin{equation}
	d(\epsilon,\alpha)\approx\frac{\sqrt{4 \epsilon -\alpha ^2}}{2 \pi  \epsilon }
\end{equation}
where only terms up to order $c^2\sim 1/M^2$ has been kept in the square root. This is equivalent to approximating the density of zeros in \eref{laguerrezerodensity} by $D(x,m)\approx\sqrt{4 M x-m^2}/(2\pi  x)$ or the density of states in the $m$ angular momentum sector by
\begin{equation}
	D(E)\approx\frac{\sqrt{2m_0 R^2 E/\hbar^2-m^2}}{2\pi E}.
	\label{EDapprox}
\end{equation}

It remains to show that this is precisely the density of states of a commutative fermion gas confined to a disk with radius $R$. For such a gas the single particle energies are given by $E_{r,m}=\hbar^2 j^2_{r,m}/(2m_0 R^2)$ where $j_{r,m}$ are the zeros of the Bessel function $J_m(j_{r,m})=0$ with  $m\in\mathbb{Z}$ and $r\in\mathbb{N}$. The zeros $\{j_{r,m}\}$ are distributed with density \cite{arriola1989}
\begin{equation}
	D_{J}(j,m)=\frac{\sqrt{1-m^2/j^2}}{\pi}
	\label{DBessel}
\end{equation}
and it is straightforward to show that this indeed produces the density of states in \eref{EDapprox}. 

\subsection{The high-density limit}
The results of the previous section suggest that to observe non-commutative effects in the thermodynamic limit we must consider densities which scale with the system size. This will render the chemical potential extensive and ensure that $c=\mut/M$ remains finite as $M$ tends to infinity. A more appropriate measure of the density is therefore $\nu\equiv \rhot/M\approx N/(2M^2)$, which will be kept fixed when taking $M\rightarrow\infty$. We also define the scaled angular momentum by $\ell\equiv L/(M^3 \hbar)$.\\

The integrals defining the $q$-potential can be approximated at low temperatures using the Sommerfeld expansion \cite{pathria}. \ref{appendixa} contains a summary of this procedure. 
The final result, up to linear order in temperature, reads
\begin{equation}
   \fl q(M,b,c,w)=\frac{M^2}{6 b}\left[\frac{c \left(b^2 c^2+\pi ^2\right) \Theta(-c)}{w^2}-\frac{(w-1)^2 \left(3 b^2 c^2+\pi ^2\right)-3 b^2 c (w-1) w^2+b^2 w^4}{(w-1)^3}\right]
   \label{q2Dexpression}
\end{equation}
where $\Theta(\cdot)$ is the step function. This result is valid for $w<1$ and $c\geq w^2/(w-1)$. The first condition is necessary to ensure that the integrals in \eref{qpot} converge. If the second condition is violated then the $f(\alpha)$ line does not intersect the $\epsilon_{\pm}(\alpha)$ curves (see \mbox{figure \ref{domainplot}}) and so $q(M,b,c,w)=0$ within this approximation.

Transforming from $(b,c,w)$ back to $(\betat,\mut,\omegat)$ yields the desired expression for $q(M,\betat,\mut,\omegat)$ in \eref{qpotdimless} in the high density, low temperature limit.
\begin{figure}[t]
\begin{center}
\begin{tabular}{cc}
\hs{-1} \includegraphics[height=7.8cm]{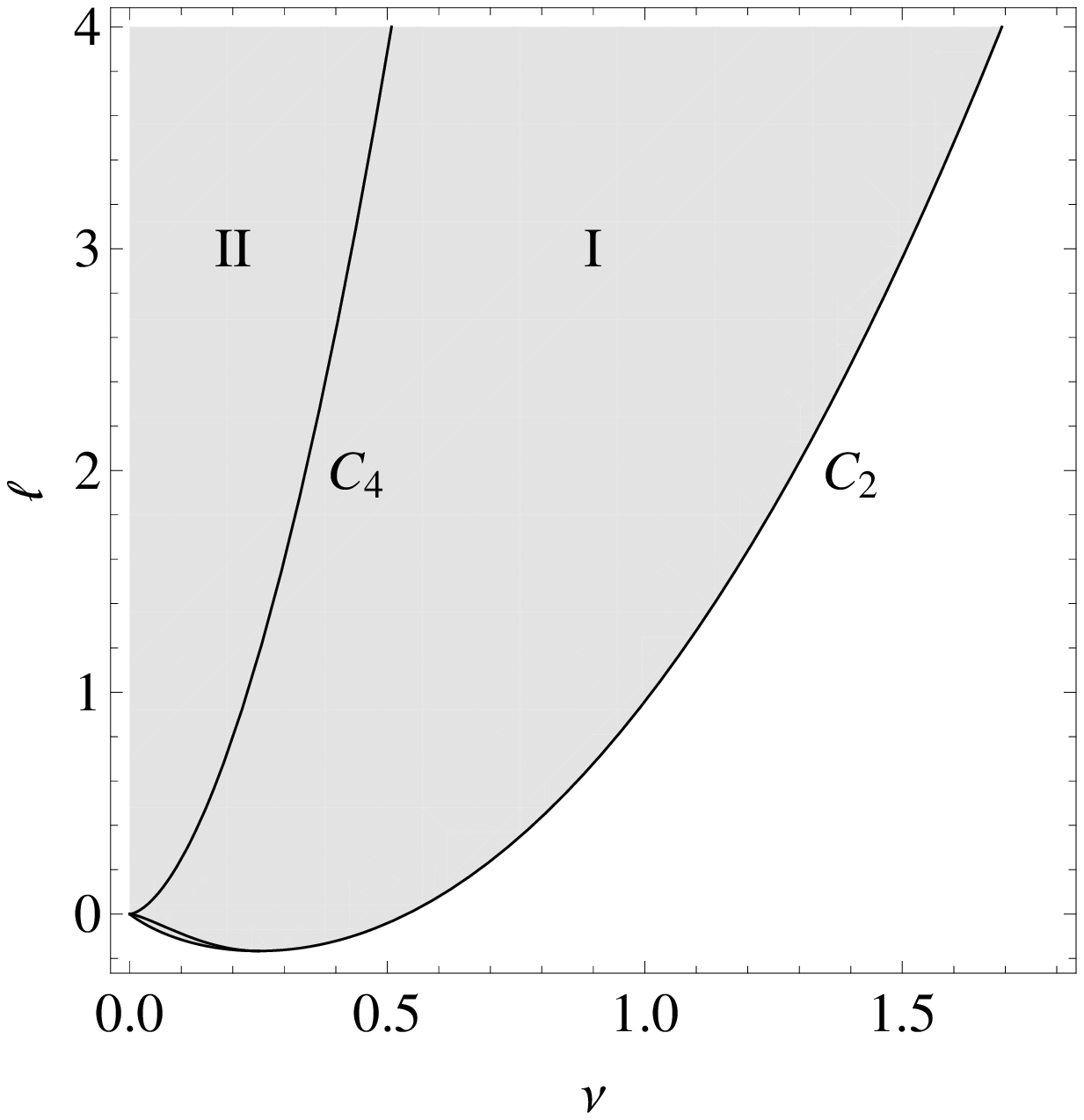} & \includegraphics[height=7.8cm,bb=0 12 370 360]{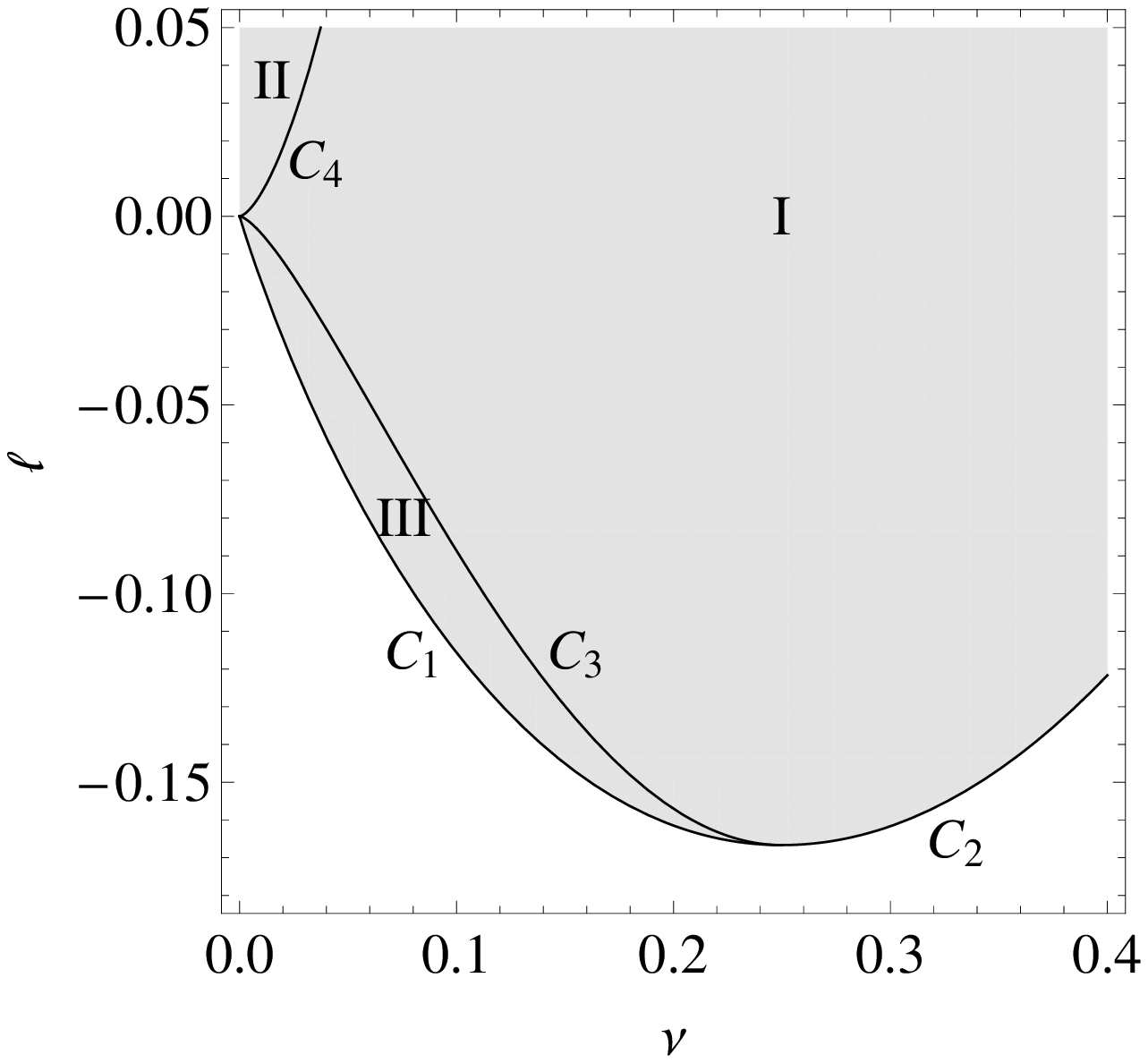}\\
(a) & (b)
\end{tabular}
\end{center}
\caption{Shaded regions indicate the physically allowed values of the scaled density $\nu=\tilde{\rho}/M$ and angular momentum $\ell=L/(M^3\hbar)$. Subfigure (b) is an enlargement of (a) around the origin.}
\label{regionplot}
\end{figure}
\subsection{Solving for $\mut$ and $\omegat$}
Our next task is to combine \eref{thermorelations} and \eref{q2Dexpression} and solve for $\mut$ and $\omegat$ as functions of the angular momentum and number of particles. It is convenient to consider the three cases (I) $\mut\geq 0$, (II) $\mut<0$, $\omegat>0$ and (III) $\mut<0$, $\omegat<0$ separately. Each of these corresponds to a physically allowed region of the $(\nu,\ell)$-plane appearing in figure \ref{regionplot}. The three regions are bounded by the curves
\numparts
\begin{eqnarray}
	C_1(\nu)=8 \nu ^{3/2}/3-2\nu \hs{2} & & C_2(\nu)=\nu  (2 \nu -1)-1/24\\
	C_3(\nu)=8 \nu ^2-16\nu ^{3/2}/3 \hs{2} & & C_4(\nu)=16\nu ^{3/2}/3+8 \nu ^2
\end{eqnarray}
\endnumparts
with $C_1(\nu)$ and $C_3(\nu)$ defined for $0\leq\nu\leq1/4$, $C_2(\nu)$ for $\nu\geq1/4$ and $C_4(\nu)$ for $\nu\geq0$. We see that the angular momentum is bounded from below by $-M^3/6$; a reflection of the finite number of negative angular momentum states. A more striking feature apparent in figure \ref{regionplot} is that for each $\ell>-1/6$ there exists a maximum density $\tilde{\rho}_c=M\nu_c$ with 
\begin{equation}
	\nu_c=(3+2\sqrt{18\ell+3})/12.
\end{equation}
No such restriction is present in the commutative case and this is a clear indication that non-commutativity results in an excluded area effect at high densities. We investigate the thermodynamic properties close to this critical density in the next section.\\

In regions (II) and (III) exact solutions for $\mut$ and $\omegat$ can be found in the thermodynamic limit. The results are
\begin{equation}
		\fl \frac{\mut}{M}=2\pm2\sqrt{\nu}-\frac{3\ell+4 \left(3\pm5 \sqrt{\nu}\right) \nu}{\left[6\nu(3\ell\pm 8 \nu ^{3/2}+6 \nu)\right]^{1/2}} \hs{1.8} \omegat=1-\left[\frac{6\nu}{3\ell\pm 8 \nu ^{3/2}+6 \nu}\right]^{1/2}
\end{equation}
where the top and bottom signs correspond to regions (II) and (III) respectively. In region (I) the exact expressions for $\mut$ and $\omegat$ are quite cumbersome and so we will focus only on the physically interesting region close to the critical density. Here we find 
\begin{equation}
	\fl \frac{\mut}{M}=1-\frac{7+24\ell-24 \nu  (6 \nu -1)}{24 \left[(8\nu +2)(\nu-\nu_0)(\nu_c -\nu)\right]^{1/2}}\hs{1.8}\omegat=1-\frac{1}{4}\left[\frac{8 \nu +2}{(\nu -\nu_0) (\nu_c-\nu )}\right]^{1/2}
	\label{muomegaatvc}
\end{equation}
where $\nu_0\equiv\left(3-2\sqrt{18\ell+3}\right)/12$. Note that as $\nu$ approaches $\nu_c$ these quantities diverge: $\mut\rightarrow\infty$ and $\omegat\rightarrow-\infty$. In figure \ref{domainplot} (a) this implies that the ``Fermi-line" $f(\alpha)=c+w\alpha$ becomes vertical at the critical density and that all states in angular momentum sectors to the left of it are occupied. 

\begin{figure}[t]
\begin{center}
\begin{tabular}{cc}
\hs{-0.8} \includegraphics[width=8cm]{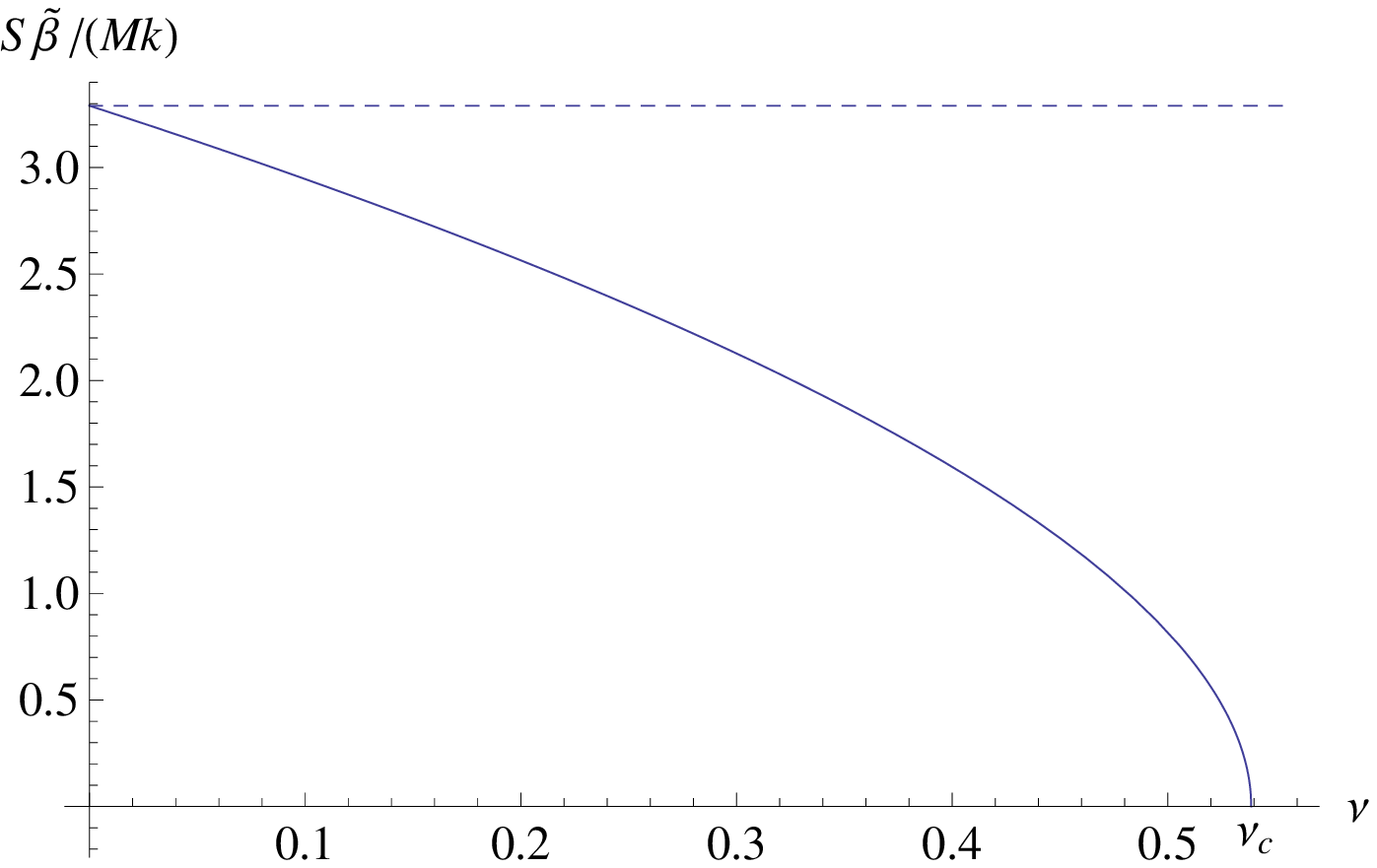} & \includegraphics[width=8cm]{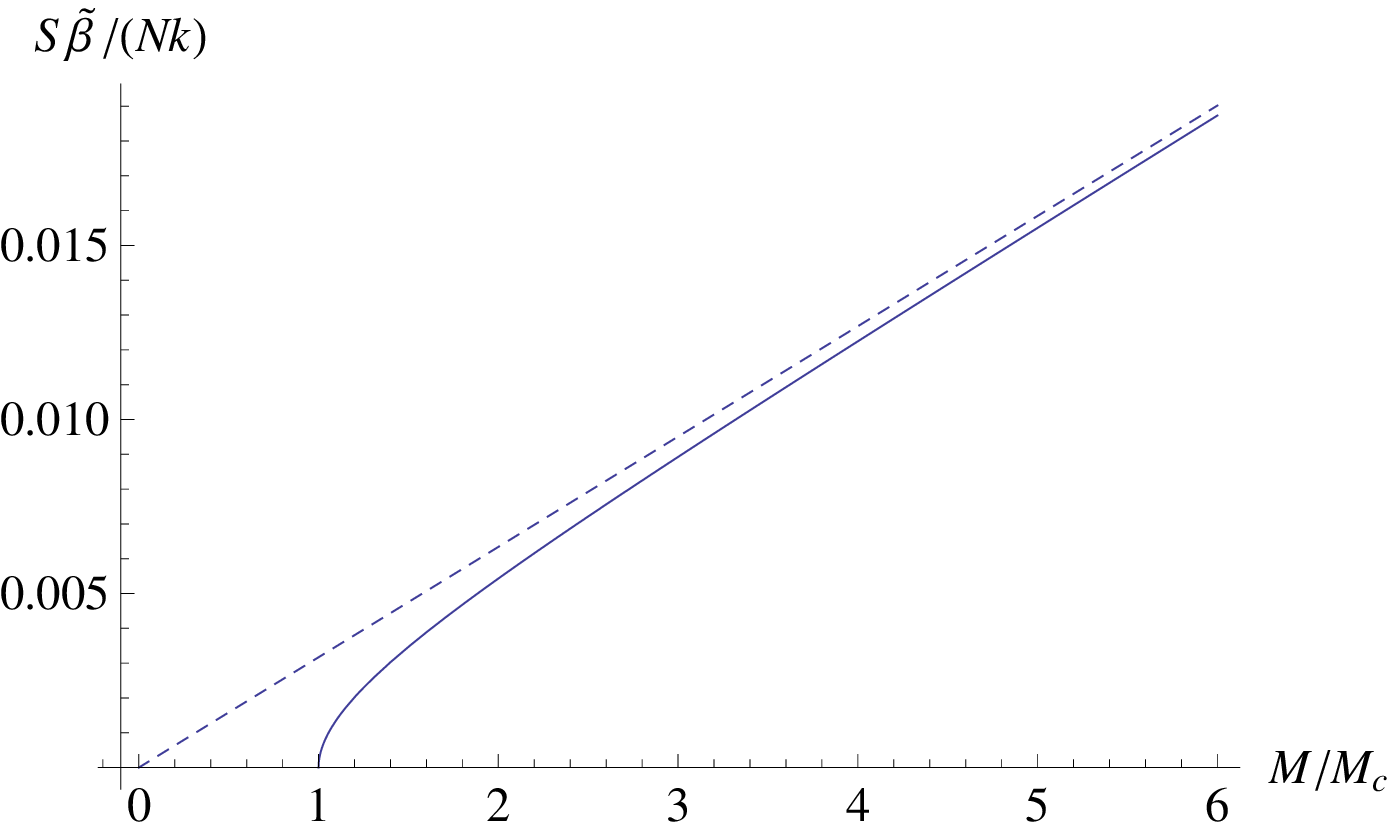}\\
(a) & (b)\\
\hs{-0.8} \includegraphics[width=8cm]{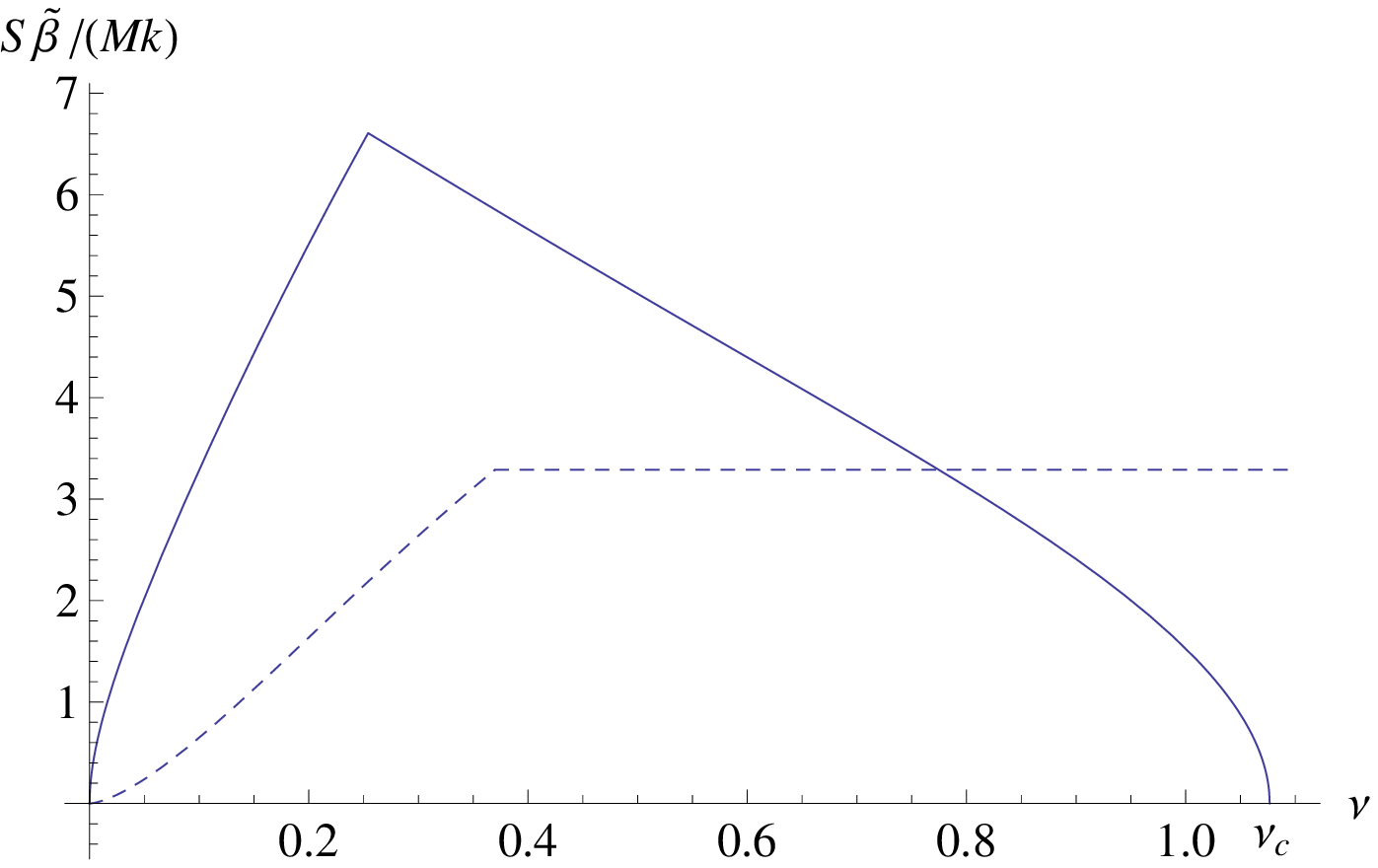} & \includegraphics[width=8cm]{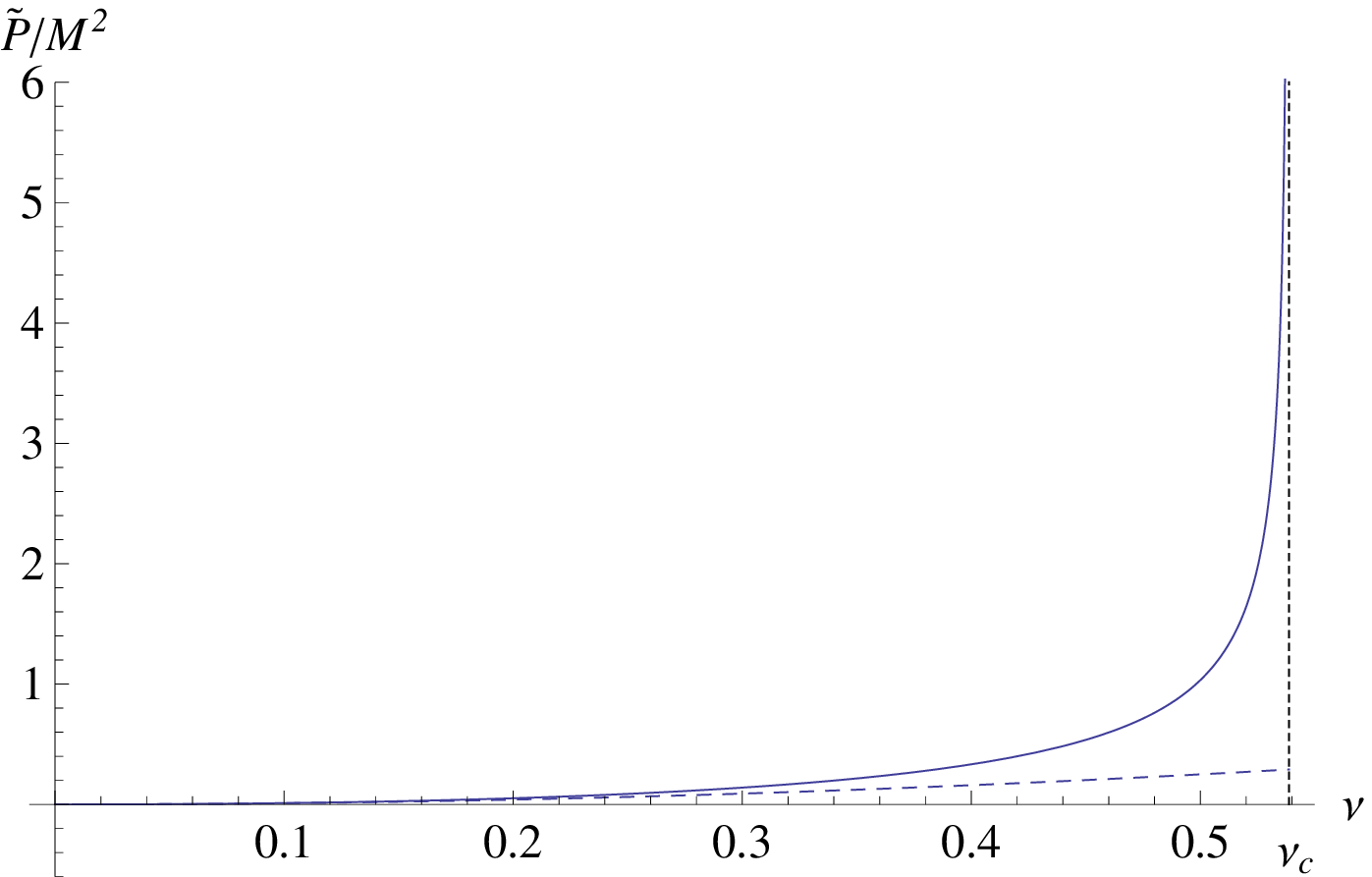}\\
(c) & (d)
\end{tabular}
\end{center}
\caption{Solid and dashed lines correspond to non-commutative and commutative results respectively. For the entropy plots $\betat$ is arbitrary but should be sufficiently large ($\betat\gtrsim100$). (a) The entropy versus $\nu$ at $\ell=0$. In the commutative case \mbox{$S/k=M\pi^2/(3\betat)$}. (b) The entropy versus $M$ for $N=10^6$ particles and $\ell=0$. Here $M_c$ is such that $\nu_c=N/(2M_c^2)$. (c) The entropy versus $\nu$ at $\ell=1.2$. (d) The pressure versus $\nu$ at $\ell=0$ for $\betat=100$. The temperature dependence of $\tilde{P}$ is however very weak.}
\label{entropy2Dplots}
\end{figure}
\subsection{Thermodynamics close to $\nu_c$}
\label{critical}
Substituting the expressions for $\mut$ and $\omegat$ in \eref{muomegaatvc} into the relations \eref{thermorelations} yields expressions for the entropy and pressure of the non-commutative gas close to the critical density:
\begin{equation}
	\fl \frac{S}{Mk}=\left[\frac{\nu_c-\nu_0}{1+4\nu_c}\right]^{1/2}\frac{2^{3/2}\pi^2}{3\betat}\sqrt{\nu_c-\nu}\hs{0.7}{\rm and}\hs{0.7}\frac{\tilde{P}}{M^2}= \left[\frac{1+4\nu_c}{\nu_c-\nu_0}\right]^{1/2}\frac{\left(1+48 \nu_c^2\right)}{96 \sqrt{2}} \frac{1}{\sqrt{\nu_c-\nu}}.
	\label{SP2d}
\end{equation}
In terms of the dimensionless system size $M$ and particle density $\tilde{\rho}\approx N/(2M)$ this implies, for example, that at $\ell=0$
\begin{equation}
	S\sim\sqrt{M}\sqrt{M \nu_c-\tilde{\rho}}\hs{0.7}{\rm  and}\hs{0.7}\tilde{P}\sim M^{5/2}/\sqrt{M \nu_c-\tilde{\rho}}.
\end{equation}
We see that as $\nu$ approaches $\nu_c$ the entropy vanishes as the square root of $\nu_c-\nu$, indicating that the ``tight packing" limit has been reached and only a single microstate is accessible to the system. This coincides with a divergence in the pressure which signals that the system is incompressible at the critical density. Due to their entropic nature these phenomenon are expected to persist at higher temperatures. In particular, the value of $\nu_c$ and the $\sqrt{\nu_c-\nu}$ dependence of the entropy and pressure should remain unaffected provided that the system is sufficiently close to its maximum density. Figure \ref{entropy2Dplots} (a) and (c) show the entropy as a function of $\nu$ at $\ell=0$ and $\ell=1.2$. The cusp in the entropy in the second case occurs at the crossing of the $C_4$ curve from region II to I. (See figure \ref{regionplot}.) In figure \ref{entropy2Dplots} (b) the number of particles is kept fixed and the system size increased from its minimum value of $M_c$. Here we again see a clear deviation from the purely extensive behaviour of the commutative case. Figure \ref{entropy2Dplots} (d) shows the pressure as a function of $\nu$ with the square-root singularity appearing at the critical density. 


The next question we address is the following: How will the entropy of systems close to the incompressible density $\nu_c$ scale with system size?  The physical motivation for this question is that we would like to know how the entropy of a very dense gas, close to its maximum density, behaves as a function of the size or number of particles when more particles are added and the size of the object increased, but in such a way that the density is always kept very close to the maximal density.  For this purpose consider a system containing $N$ particles with a minimum size $M_c$ given by $M_c^2=N/(2\nu_c)$. The density can then be expressed as  $\nu=N/(2M^2)=\nu_c M_c^2/M^2$. The limit that interest us is when the system size exceeds the minimum size by a small fixed area $2\pi\theta\Delta M$. Then $M=M_c+\Delta M$ with  $\Delta M$ a non-negative integer such that $\Delta M<<M_c$ and $\nu\approx\nu_c(1-2\Delta M/M_c)$, i.e., the density is close to the incompressible density. Crucial here is that the quantization of the system size implies that $\Delta M\sim 1$. With $\Delta M$ held fixed we now consider the dependence of the entropy on the system size $M\approx M_c$. Substituting in \eref{SP2d} the entropy close to the incompressible point reads
\begin{equation}
S(M_c)\sim M\sqrt{\nu_c}\sqrt{1-\frac{M_c^2}{M^2}}\approx \sqrt{2\nu_c\Delta M}\sqrt{M_c}.
\end{equation}
As $R^2\approx 2M\theta$ it follows that the entropy is proportional to the circumference, rather than the area, of the system.  When $M>>M_c$, i.e. far from the incompressible point, normal extensive behaviour is recovered. 

Interestingly, the proportionality constant takes discrete values ($\Delta M=0,1,2\ldots$). 
The way to interpret this is probably to think in terms of families of systems parameterized by the number of particles or the minimal system size.  Each family is characterized by a specific value of $\Delta M$, i.e., for minimal size $\Delta M=0$ the entropy vanishes, for $\Delta M=1$ the entropy behaves as $S(M_c)\sim\sqrt{2\nu_c}\sqrt{M_c}$ etc.  Within the current context it is, of course, not possible to determine the value of this proportionality constant.  

This is a rather remarkable result, especially since it follows from a first principle calculation, 
as it implies that the number of microscopic degrees of freedom for such dense systems is actually determined by the circumference rather than the area (size) of the system.  
It is also worthwhile noting that the argument that leads to this result has only a number of generic features and essentially relies only on the existence of an incompressible point and the quantization of system size.  This robustness of this result can also be understood from an intuitive point of view.  In essence the minimum size corresponds to ``tight packing" for which only one microstate is possible and thus the entropy vanishes (see figure 1).  For values of the system size slightly above than the minimal size, and thus for densities slightly lower than the incompressible density, only a fraction of the particles in a thin shell around the Fermi level determine the thermodynamic behaviour; the bulk of particles is still frozen in a ``tightly packed core".  Furthermore, the particles in this thin shell occupy the highest single particle angular momentum states and are therefore expected to be close to the edge of the well.  From this the intuitive picture is therefore that in real space a thin shell of particles at the edge of the well determines the thermodynamic properties of the system.  Indeed, assuming a shell thickness $\Delta R$ and that each particle occupies an area $\theta$, one easily concludes that the number of microstates behave as $\Omega\sim 2^{2\pi R\Delta R/\theta}$ and thus that the entropy behaves as $S\sim R$ with the proportionality constant set by the shell thickness as was also found above. From this perspective the behaviour found above is not surprising and expected to be rather generic.

\section{The three-dimensional non-commutative fermion gas}
\label{threedim}
We assume the following commutation relations in three dimensions
\begin{equation}
\label{nc3d}
[x_i,x_j]=i\theta_{i,j}
\end{equation}
with $\theta_{ij}$ a real anti-symmetric matrix.  Through an appropriate choice of coordinates it is always possible to restrict the non-commutativity to two of the spacial coordinates with the third coordinate being commutative. We consider a cylindrical geometry with radius $R$ and length $L$ with its central axis orientated along the $z$-direction. The $x$- and $y$-directions are taken to be non-commutative and at a fixed $z$ the problem therefore reduces to the two-dimensional case considered earlier. In the $z$-direction we have the regular one-dimensional free particle problem with vanishing boundary conditions. In the notation of \eref{qpotdimless} the $q$-potential reads
\begin{eqnarray}
	q_{3D}(M,\betat,\mut,\omegat,\gammat)&=&\sum_{n=1}^\infty\sum_{m,r}\log[1+e^{-\betat (x_{r,m}+\gammat n^2-\mut-\omegat m)}]\\
	&=&\sum_{n=1}^\infty q_{2D}(M,\betat,\mut-\gammat n^2,\omegat)
	\label{qpotdimless3D}
\end{eqnarray}
where $\gammat=\gamma/E_0$ and $\gamma=\hbar^2\pi^2/(2m_0 L^2)$. In the thermodynamic limit $R$ and $L$ will tend to infinity in a fixed ratio. This can be made explicit by writing $\gammat=G/M$ where $G=(\pi/2)^2 (R/L)^2$ is a dimensionless constant defining the shape of the cylinder. As is evident from \eref{qpotdimless3D} the $q_{3D}$-potential is just a sum of two-dimensional potentials with shifted chemical potentials. The sum terminates automatically once $\mut-\gammat n^2$ drops below $M \omegat^2/(\omegat-1)$, as discussed after equation \eref{q2Dexpression}. Under the same assumptions that led to $q_{2D}$ in \eref{q2Dexpression} we can now express $q_{3D}$ as 
\begin{equation}
	q_{3D}(M,\betat,\mut,\omegat,\gammat)=\int_0^{n_+} {\rm d}n\ q_{2D}(M,\betat,\mut-\gammat n^2,\omegat)
	\label{qpot3D}
\end{equation}
where $n_+>0$ is such that $\mut-\gammat n_+^2=M\omegat^2/(\omegat-1)$. This integral is straightforward to perform since $q_{2D}$ is a piecewise polynomial in its third argument. 
\begin{figure}[t]
\begin{center}
\begin{tabular}{cc}
\hs{-0.6} \includegraphics[width=7.8cm]{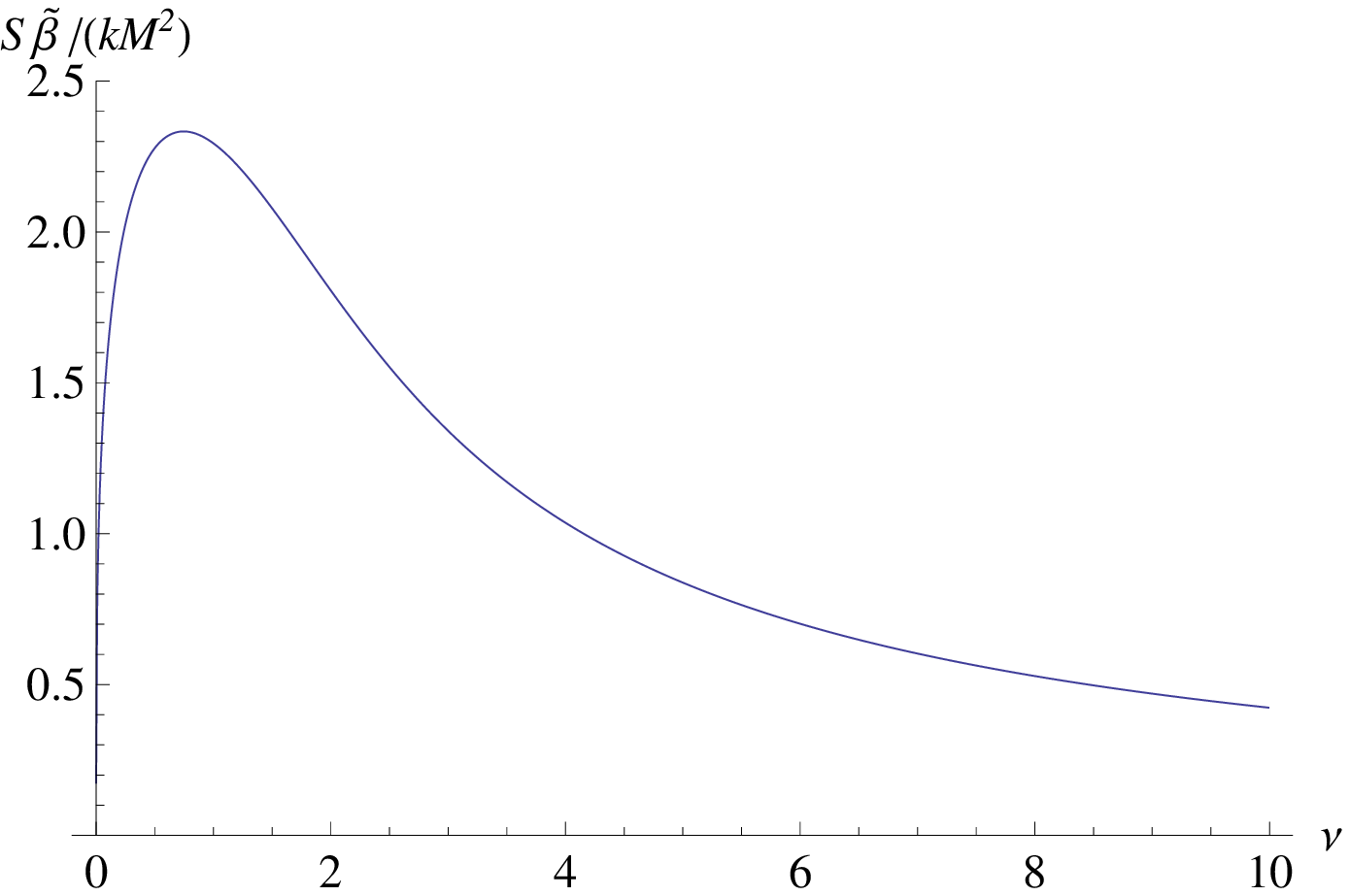} & \includegraphics[width=7.8cm]{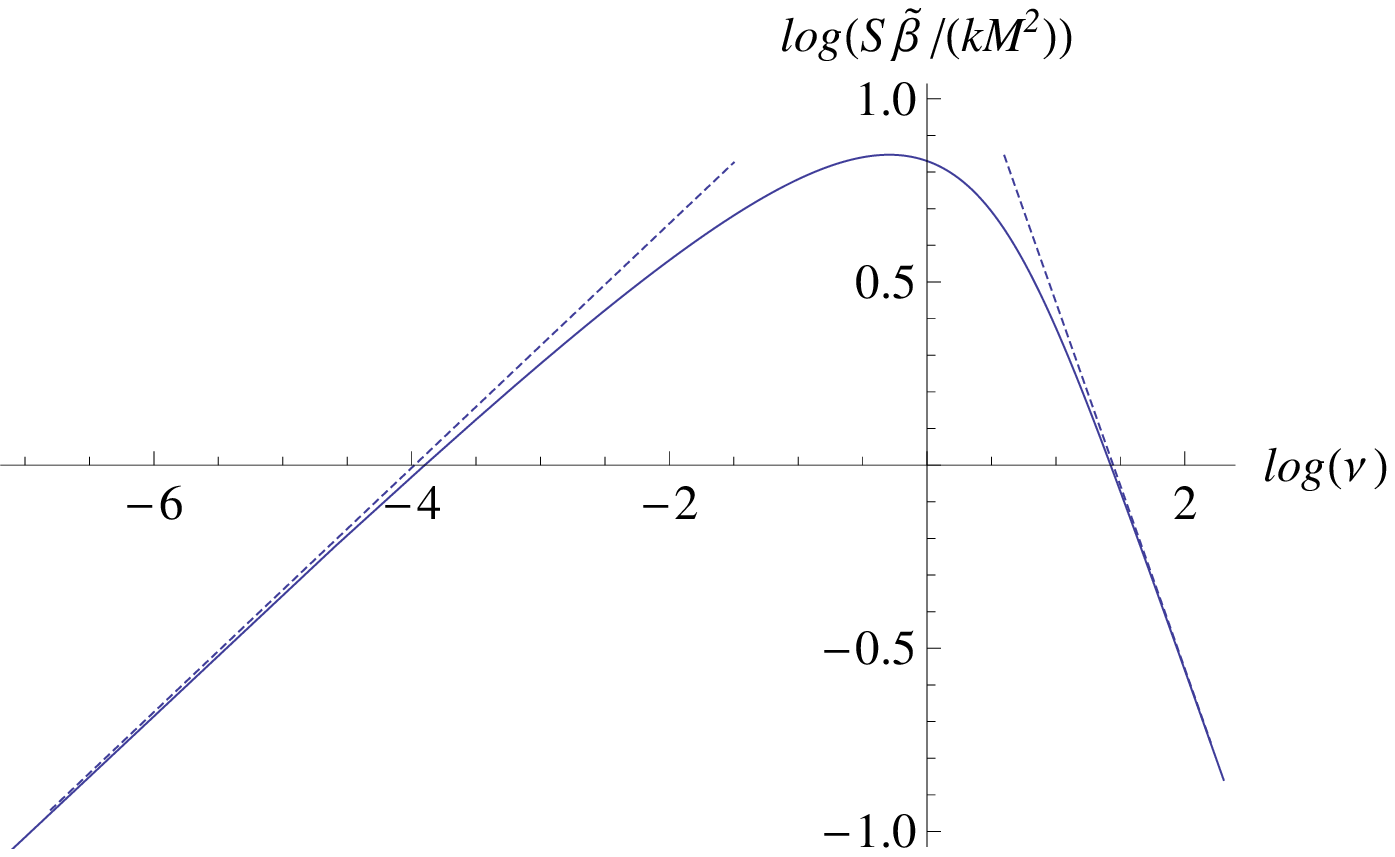}\\
(a) & (b)
\end{tabular}
\end{center}
\caption{(a) The entropy versus $\nu$ at $\ell=0$. (b) A log-log plot of the entropy versus $\nu$. The dashed lines indicate the high and low density behaviour and are based on the expressions in \eref{S3Dexpressions}. }
\label{entropy3Dplots}
\end{figure}
\subsection{Scaling behaviour of the entropy}
In three dimensions it is convenient to use $\nu=N/M^3$ and $\ell=L/(\hbar M^4)$ as measures of the density and angular momentum. Combining \eref{qpot3D} and \eref{thermorelations} leads to a set of equations for $\mut$ and $\omegat$ which can be solved numerically for given $\nu$ and $\ell$. This produces the results of figure \ref{entropy3Dplots}. In (a) we show the entropy as a function of $\nu$ for $\ell=0$. Unlike the two-dimensional case there is no maximum density at which the entropy vanishes and pressure diverges, i.e. no incompressible point. This reflects the fact that the excluded \emph{area} effect is restricted to the $x-y$ plane and does not place any restriction on how the commutative modes in the $z$-direction are occupied. However, the effects of non-commutativity are still present and responsible for modifying the scaling behaviour of the entropy as we progress from low to high densities. This is clearly reflected in figure \ref{entropy3Dplots} (b) which reveals that $S$ grows like $\nu^{1/3}$ at low densities and decreases like $1/\nu$ at high densities. In these two limits we are able to solve for $\mut$ and $\omegat$ when $\ell=0$ to find
\numparts
\begin{eqnarray}
	\hs{-1}\mut=M\left(9 G\nu^2/4\right)^{1/3} \hs{1.5} && \omegat=-\left(18 G\nu^2\right)^{1/3}/5\hs{1.61} (\nu<<1) \label{muomegald}\\
	\hs{-1}\mut=M(0.72703\ldots) G\nu^2 \hs{1.5} && \omegat=-(0.82342\ldots) G\nu^2\hs{1.2} (\nu>>1) \label{muomegahd}
\end{eqnarray}
\endnumparts
The constants in \eref{muomegahd} are derived from the numerical solution of a transcendental equation. The entropy is
\begin{equation}
	\frac{S}{kM^2}=\left[\frac{\pi^6}{18 G\betat^3}\right]^{1/3} \nu^{1/3}\hs{1}{\rm and}\hs{1}\frac{S}{kM^2}=\frac{(4.2397\ldots)}{G \betat}\frac{1}{\nu}
	\label{S3Dexpressions}
\end{equation}
for low and high densities respectively. These results provide the dashed lines in the log-log plot in figure \ref{entropy3Dplots} (b). In terms of the dimensionless volume $\tilde{V}=V/\theta^{3/2}$ and density $\tilde{\rho}=N/\tilde{V}$ these expressions become
\begin{equation}
	\frac{S}{k}=\frac{1}{\betat}\left(\frac{\pi }{6}\right)^{2/3}\tilde{V}\tilde{\rho}^{1/3}\hs{1}{\rm and}\hs{1}\frac{S}{k}=\frac{\eta G^{2/3}}{\betat}\frac{\tilde{V}^{7/3}}{\tilde{\rho}}.
\end{equation}
with $\eta\approx6.4757\times10^{-4}$. The low density result agrees exactly with that of a commutative fermion gas \cite{pathria}. It is independent of both the shape of the cylinder (i.e. $G$) as well as the non-commutative parameter $\theta$. In contrast, the high density result depends on both these parameters and also exhibits an unusual non-extensive volume dependence. Furthermore, this entropy is inversely proportional to the particle density as is the case for a one-dimensional ideal gas. This observation that some aspects of the high density thermodynamics mimic that of a one-dimensional system is also observed in the relation between pressure and energy density. For a non-interacting gas in $d$ dimensions it holds quite generally that $P=(2/d)(E/V)$. We indeed find that $P=(2/3)(E/V)$ and $P=2(E/V)$ in the low and high densities limits respectively. 

These results hint that this type of non-commutativity and constraint on $J_z$ lead to a high density limit that behaves as a commutative one-dimensional system.  This counters our expectations for the real world and we can therefore only assume that this form of non-commutativity, the symmetry breaking it implies and the constraint on $J_z$ are inappropriate.  In particular the dependence on the shape of the container, i.e., the $G$ dependence above is apparently due to the breaking of rotational symmetry.  This is confirmed by comparing with the commutative case where such a dependence also occurs due to the $J_z$ constraint we imposed here, i.e. due to the breaking of rotational symmetry.  This suggests that the restoration of rotational symmetry and the imposition of the more appropriate constraint on $J$ are crucial for recovering the appropriate high density behaviour.  In this regard it is important to remark that we do not expect that the restoration of the symmetry through twisting and the corresponding issue of twisted statistics alone can lead to the appropriate high density thermodynamics, but that the constraint on $J$ is also a key ingredient. The reason for this is that the twist simply implies a unitary transformation implemented through the twist operator \cite{bal} and thus should not affect the thermodynamics as the computation of the partition function involves a trace.  However, correlation functions are affected by twisting as was shown in \cite{chak1}. 

Attempts at restoring the rotational symmetry through twisting and constraining $J$ may also be problematic as it was shown in \cite{chak2,sch2} that even though twisting restores the vectorial transformation properties of the non-commutative coordinates, the symmetry is still not manifest on the level of the Hamiltonian and Schroedinger equation, i.e., the Hamiltonian does not commute with $J^2$. It is therefore not clear that this constraint can be imposed consistently in a statistical ensemble setting.  

A natural alternative is to change the non-commutative relations from the simple form (\ref{nc3d}) to those of a fuzzy sphere $[x_i,x_j]=i\theta\epsilon_{ijk}x_k$.  In this way $R^3$ can be realized as an ``onion structure" with quantized radius and the two-dimensional program of \cite{scholtz2008} can in principle be executed to compute the spectrum of a three-dimensional well. Unfortunately this program is technically extremely demanding and has not yet been performed successfully.  Some progress in this direction was made in \cite{shariati} and seems to confirm the expectations described here.

\section{Conclusions}
\label{concl}
The thermodynamic behaviour of two- and three-dimensional non-commutative gases was investigated analytically in the high density region. Strong deviations from conventional commutative behaviour was found. In particular an incompressible point was found in two dimensions and the entropy exhibits in both cases non-extensive behaviour.  More profoundly, for two-dimensional systems with density close to the incompressible density the entropy exhibits a very particular dependence on the system size in that it is proportional to the square root of the system size, i.e. the circumference rather than the area. Incompressibility is absent in three dimensions and the scaling of entropy at high density is unconventional, showing rather a crossover to one-dimensional commutative thermodynamics.  This is probably due to the particular form of non-commutativity used here in three dimensions and the symmetry breaking it implies. In particular the constraint on $J_z$, rather than $J$, is problematic and cannot be implemented correctly in the current framework.  Rectifying this situation through a twisting of the symmetry seems problematic and a revision of the non-commutative relations themselves may be required.    

It will also be interesting to study, even in two dimensions, non-commutative fermion gases confined by a mutual attractive interaction between particles rather than a confining potential.  In this case an interesting interplay between the excluded volume, thermal length and interaction length scales may be expected.  

\section{Acknowledgements}
This work was supported under a grant of the National Research Foundation of South Africa. 

\newpage
\appendix
\section{Calculating the $q$-potential}
\label{appendixa}
Our goal is to derive an approximate expression for the 2D $q$-potential
\begin{equation}
	q(M,b,c,w)=M^2\int_{-1}^{\infty}{\rm d}\alpha\int_{\epsmin}^{\epsplus}{\rm d}\epsilon\ d(\epsilon,\alpha)\log[1+e^{-b (\epsilon-f(\alpha))}] 
	\label{qpotappendix}
\end{equation}
at low temperatures and high densities. Here $f(\alpha)=c+w\alpha$ and
\begin{equation}
	d(\epsilon,\alpha)=\frac{\sqrt{4 \epsilon -(\alpha -\epsilon )^2}}{2\pi\epsilon}=\frac{\sqrt{(\epp-\epsilon)(\epsilon-\epm)}}{2\pi\epsilon}
\end{equation}
with 
\begin{equation}
	\epsilon_{\pm}(\alpha)=\alpha+2 \pm 2 \sqrt{\alpha +1}.
\end{equation}
The values of $\alpha$ at which $f(\alpha)$ intersects the $\epsilon_{\pm}(\alpha)$ curves are
\begin{equation}
	\alpha_\pm=\frac{c-cw+2w\pm2\sqrt{c-c w+w^2}}{(1-w)^2}.
\end{equation}
See figure \ref{domainplot}. Let $\mathcal{I}(\alpha)$ denote the $\epsilon$-integral in \eref{qpotappendix} for a fixed $\alpha$. An expression for $\mathcal{I}(\alpha)$ can be derived in terms of the three functions
\begin{eqnarray}
\fl I(\epsilon,\alpha)&=&\int_\epm^\epsilon d(\epsilon',\alpha){\rm d}\epsilon'=\epsilon  d(\epsilon,\alpha)+\alpha\Theta(-\alpha)+\frac{1}{\pi}\arccos\left[\frac{\epm+\epp-2 \epsilon}{\epp-\epm}\right]\nonumber \\
\fl & &\hs{2.8}-\frac{\epm+\epp-4}{2\pi}\arccos\left[\frac{\epm+\epp+2 \epsilon -4}{(\epp-\epm)\sqrt{\epsilon }}\right]\nonumber \\
\fl D(\epsilon,\alpha)&=&\int_\epm^\epsilon \epsilon'd(\epsilon',\alpha){\rm d}\epsilon'=\frac{(\epm-\epp)^2}{16\pi}\arccos\left[\frac{\epm+\epp-2 \epsilon}{\epp-\epm}\right]-\frac{(\epm+\epp-2 \epsilon)}{4}\epsilon d(\epsilon,\alpha)\nonumber \\
\fl L(\epsilon,\alpha)&=&\int_\epm^\epsilon I(\epsilon',\alpha){\rm d}\epsilon'=\epsilon I(\epsilon)-D(\epsilon)\nonumber
\end{eqnarray}
Here $\epsilon_{\pm}\equiv\epsilon_{\pm}(\alpha)$. Important special cases are $I(\epm,\alpha)=0$, $I(\epp,\alpha)=1+\alpha\Theta(-\alpha)$, $D(\epm,\alpha)=0$ and $D(\epp,\alpha)=(\epp-\epm)^2/16$.\\

To evaluate $\mathcal{I}(\alpha)$ at low $T$ (large $b$) there are three cases to consider:
\begin{enumerate}
	\item $f(\alpha)>\epp$: Here $-b(\epsilon-f(\alpha))>0$ and so $\exp[-b (\epsilon-f(\alpha))]$ is exponentially large in $b$. We approximate 
\begin{equation}
	\log[1+e^{-b (\epsilon-f(\alpha))}]\approx -b (\epsilon-f(\alpha))
\end{equation}
to produce
	\begin{equation}
		\fl \mathcal{I}(\alpha)\approx b f(\alpha) I(\epp,\alpha)-b D(\epp,\alpha)=bf(\alpha)(1+\alpha \Theta(-\alpha))-\frac{1}{16} b (\epp-\epm)^2
	\end{equation}
	In figure \ref{domainplot} (a) this case corresponds to the region $-1\leq\alpha\leq\alpha_-$. 
	\item $\epm\leq f(\alpha)\leq\epp$: Here the effective chemical potential $f(\alpha)$ lies inside the integration domain of the $\epsilon$ integral and we use the Sommerfeld expansion \cite{pathria} to approximate the integral up to linear order in $b^{-1}$. Integrating by parts twice and then expanding the integrand around $\epsilon=f(\alpha)$ produces 
	\begin{equation}
	\mathcal{I}(\alpha)\approx bL(f(\alpha),\alpha)+\frac{\pi^2}{6 b} d(f(\alpha),\alpha).
	\end{equation}
	In figure \ref{domainplot} (a) this case corresponds to the region $\alpha_-\leq\alpha\leq\alpha_+$.
	\item $f(\alpha)<\epm$: Here $-b(\epsilon-f(\alpha))<0$ and the integrand in exponentially small in $b$ for all $\epsilon\in[\epm,\epp]$. We therefore approximate $\mathcal{I}(\alpha)\approx0$. In figure \ref{domainplot} (a) this case corresponds to the region $\alpha>\alpha_+$.
\end{enumerate}

With $\mathcal{I}(\alpha)$ known it remains to perform the $\alpha$-integral in \eref{qpotappendix}. This involves extensive algebraic manipulations which we will not present here. The integrals themselves are however standard and may be found in texts such as \cite{ryzhik2007}.

\section*{References}
\bibliographystyle{plain}

\begin{thebibliography}{10}
\bibitem{dop} Doplicher S, Fredenhagen K and Roberts J E 1995 {\it Commun. Math. Phys.} {\bf 172} 187
\bibitem{doug} Douglas M R and Nekrasov N A 2001 {\it Rev. Mod. Phys.} {\bf 73} 97
\bibitem{scholtz2009} Scholtz F G, Gouba L, Hafver A and Rohwer C M 2009 {\it Jnl. Phys. A} {\bf 42}  175303
\bibitem{li} Li K and Dulat S 2006 {\it Eur. Phys. J. C} {\bf 46} 825
\bibitem{mendes} Vilela Mendesa R 2005 {\it Eur. Phys. J. C} {\bf 42} 445
\bibitem{bem} Bemfica F S and Girotti H O 2005 {\it Jnl. Phys. A} {\bf 38} L539
\bibitem{khan} Khan S, Chakraborty B and Scholtz F G 2008 {\it Phys. Rev. D} {\bf 78} 025024
\bibitem{chai} Chaichian M, Sheikh-Jabbari M M and Tureanu A 2001 {\it Phys. Rev. Lett.} {\bf 86} 2716
\bibitem{chair} Chair N and Sheikh-Jabbari M M 2001 {\it Phys. Lett. B} {\bf 504} 141
\bibitem{lia} Liaoa Y and Dehneb C 2003 {\it Eur. Phys. J. C} {\bf 29} 125
\bibitem{ohl} Ohl T and Reuter J 2004 {\it Phys. rev. D} {\bf 70} 076007
\bibitem{gar} Garc\'{i}a-Compe\'{a}n H, Obreg\'{o}n O and Ram\'{i}rez C 2002 {\it Phys. Rev. Lett.} {\bf 88} 161301
\bibitem{chai1} Chaichian M, Kulish P P, Nishijima K and Tureanu A 2004 {\it Phys. Lett. B} {\bf 604} 98
\bibitem{alex} Alexander S, Brandenberger R and Magueijo J 2003 {\it Phys. Rev. D} {\bf 67} 081301
\bibitem{scholtz2007} Scholtz F G, Chakraborty B, Govaerts J and Vaidya S 2007 {\it Jnl. Phys. A} {\bf 40} 14581
\bibitem{wess1} Wess J 2004 Deformed coordinate spaces: Derivatives arXiv:hep-th/0408080
\bibitem{asch} Aschieri P, Blohmann C, Dimitrijevic M, Meyer F, Schupp P and Wess J 2005 {\it Class. Quant. Grav.} {\bf 22} 3511
\bibitem{bal} Balachandran A P, Mangano G, Pinzul A and Vaidya S 2006 {\it Int. J. Mod. Phys. A} {\bf 21} 3111
\bibitem{chak1} Chakraborty B, Gangopadhyay S, Hazra A G and Scholtz F G 2006 {\it Jnl. Phys. A} {\bf 39} 9557
\bibitem{wess} Fiore G and Wess J 2007 {\it Phys. Rev. D} {\bf 75} 105022 
\bibitem{fiore} Fiore G 2008 On the consequences of twisted Poincare symmetry upon QFT on Moyal noncommutative spaces arXiv:hep-th/0809.4507
\bibitem{scholtz2008}  Scholtz F G and Govaerts J 2008 {\it Jnl. Phys. A}  {\bf 41} 505003
\bibitem{arriola1989} Arriola E R and Dehesa J S 1989 {\it Nuovo Cimento B} {\bf 103} 611
\bibitem{dette1995} Dette H and Studden W J 1995 {\it Constructive Approximation} {\bf 11} 227
\bibitem{pathria} Pathria R K 1972 {\it Statistical Mechanics} (Oxford: Pergamon)
\bibitem{shariati}  Shariati A, Khorrami M and Fatollahi A H 2010 {\it Jnl. Phys. A}  {\bf 43} 285001
\bibitem{ryzhik2007} Gradshteyn I S and Ryzhik I M 2007 {\it Table of integrals, series and products} 7th ed (Burlington: Academic Press)
\bibitem{chak2} Chakraborty B, Kuznetsova Z and Toppan F 2010 {\it J.Maths.Phys.} {\bf 51} 112102
\bibitem{sch2} Sinha D, Chakraborty B and Scholtz F G, "Non-commutative Quantum Mechanics in Three Dimensions and
Rotational Symmetry", arXiv:hep-th/1108.2569 
\end{thebibliography}

\end{document}